\documentclass[12pt]{article}
\begin{document}
\def\pd{\partial}
\section*{Another Complex Bateman  Equation}

{D. B. {Fairlie}\footnote{e-mail: {david.fairlie@durham.ac.uk}}}\\
{ Department of Mathematical Sciences\\
University of Durham, Durham DH1 3LE}

\begin{abstract}A further class of complex covariant field equations is investigated. 
These equations possess several common features: they may be solved, or partially solved in terms of implicit functional relations, they possess an infinite number of inequivalent 
Lagrangians which vanish on the space of solutions of the equations of motion, they are invariant under linear transformations of the independent variables, and thus are signature-blind and are consequences of first order equations of hydrodynamic type. 
\end{abstract}

\section{Introduction}
This paper is another in a series devoted to an investigation of simple equations exhibiting covariance of solutions. These equations have arisen in the study of generalisations of the Bateman equation \cite{gov}, in the equations arising from continuations of the String and Brane Lagrangians to the situation where the target space has fewer dimensions than the base space \cite{baker} and a complex form of these equations
\cite{fai}. The simplest example of these is a complexification of the Bateman equation.
What we have called the Complex Bateman equation is the following equation for a real function $\phi$ defined over the space of variables $(x_1,\ x_2;\ \bar x_1,\ \bar x_2)$;
\begin{equation}
\phi_{x_1}\phi_{\bar x_1}\phi_{x_2\bar x_2} + \phi_{x_2}\phi_{\bar x_2}\phi_{x_1\bar x_1}-\phi_{x_1}\phi_{\bar x_2}\phi_{\bar x_1 x_2}-\phi_{x_2}\phi_{\bar x_1}\phi_{x_1\bar x_2}\,=\,0.\label{complex1}
\end{equation}
(Here subscripts denote differentiation). This equation was shown to be
completely integrable \cite{chaundy}\cite{leznov}, with solution given by
solving for $\phi$ the following constraint upon two arbitrary functions of three variables, $F,\ G$;
\[ F(\phi;\, x_1,x_2)\,=\,G(\phi;\,\bar x_1,\bar x_2).\]
From the form of the solution, or from the equation itself, it is manifest that if $\phi$ is a solution, any function of $\phi$ will also be a solution and thus that the equation exhibits covariance. It is also invariant under separate diffeomorphisms of the pairs of variables $(x_1,\ x_2)$ and $(\bar x_1,\ \bar x_2)$. In fact a subclass of solutions is given by the sum of `holomorphic'  and `antiholomorphic' functions
\[\phi\,=\,f(x_1,x_2) +g(\bar x_1,\bar x_2).\]  
A general characteristic of such equations is that they possess an infinite number of inequivalent Lagrangians. The equations of motion are partially, or sometimes fully solveable in implicit form, as in the examples cited. The fully integrable eqautions arise from kinematical first order equations of hydrodynamic type.

\section{ Another Complex Bateman Equation}

Now there is another possibility for complexification; we could take instead
\begin{equation}
\bar\phi_x \phi_x\phi_{tt}-\bar\phi_x \phi_t\phi_{tx}-\bar\phi_t \phi_x\phi_{tx}+\bar\phi_t\phi_t\phi_{xx}\,=\,0,\label{complex2}
\end{equation}
together with its complex conjugate. These equations also exhibit covariance; $\phi$ may be replaced by any function of itself, and the same for $\bar\psi$ and the equations remain invariant.
Where do these equations come from?
Take the hydrodynamic equations
\begin{eqnarray}
\frac{\partial u}{\partial t} &=&v\frac{\partial u}{\partial x},\label{int1}\\
\frac{\partial v}{\partial t} &=&u\frac{\partial v}{\partial x},\label{int2}
\end{eqnarray}
and set $\displaystyle{u\,=\, \frac{\bar\phi_t}{\bar\phi_x},\ v\,=\, \frac{\phi_t}{\phi_x}}$, and the equation (\ref{complex2}), together with its complex conjugate are reproduced. Indeed, all that is necessary is to set
in an alternative reduction, $u=\bar\phi$ and $v=\phi$ and the same equations arise in consequence. 
These equations admit an infinite number of conserved quantities \cite{mulvey};
If $S_n$ denotes the symmetric polynomial of degree $n$ in $u,\ v$, then
\begin{equation}
\frac{\pd }{\pd t}S_n \,=\,\frac{\pd }{\pd x}(uv S_{n-1})\label{sym}
\end{equation} is a conservation law. This is easily proved by induction and from the
iterative definition: $S_n\,=\, u^n+vS_{n-1}$.                  
These equations can be integrated by the usual hodographic method of interchanging dependent and independent variables, where they become
\begin{eqnarray}
\frac{\partial x}{\partial v} &+&v\frac{\partial t}{\partial v}\,=\,0,\label{inv1}\\
\frac{\partial x}{\partial u} &+&u\frac{\partial t}{\partial u}\,=\,0,\label{inv2}
\end{eqnarray}
which can be solved in terms of two arbitrary functions $f,\ g$ to give
\[ t\,=\,f'(u)+g'(v);\ \ \ x\,=\,f(u)-uf'(u)+g(v)-vg'(v)\]
with primes denoting differentiation with respect to the argument. If $u=\bar\phi$ and $v=\phi$ this parametrisation is a  solution to the alternative complexification. Note that the requirements that $(t,\ x)$ be real imposes a further constraint upon the functions $(f,\ g)$. Of course, if $(\phi,\ \bar\phi)$ are treated as independent real functions, no such restriction exists. Now the second order equations (\ref{complex2}) are Poincar\'e invariant; indeed are covariant under general inhomogeneous linear transformations of the independent variables.
 This must be true also for the first order equations (\ref{inv1}),
(\ref{inv2}). They are clearly translation invariant; if $(t,\ x)$ transform as
\[t'\,=\,at+bx,\\\ x'\,=\,ct+dx,\]
then invariance will be maintained if 
\[u'\, =\, \frac{du-c}{a-bu},\ \ \ \ v'\, =\, \frac{dv-c}{a-bv}.\]
\section{2-dimensional Born-Infeld equation}
We may remark parenthetically that the same equations(\ref{int1}),(\ref{int2}) also yield the general solution to one form of the so-called Born-Infeld equation in two dimensions in light-cone co-ordinates \cite{mulvey}\cite{arik};
\[
\phi_x^2 \phi_{tt}+\phi_x^2 \phi_{tt}-(\lambda+2\phi_x\phi_t)\phi_{xt}\,=\,0.
\]
This is achieved by setting
\[ \frac{\pd\phi}{\pd x}\,=\,\frac{\sqrt{\lambda}}{\sqrt{u}-\sqrt{v}},\ \ \ \frac{\pd\phi}{\pd t}\,=\,\frac{\sqrt{\lambda uv}}{\sqrt{u}-\sqrt{v}}.\]
The integrability constraints upon these equations is just the Born-Infeld equation itself. 
Thus the primacy of the first order hydrodynamic equations is again manifest. This is a phenomenon which has been noticed before; that the same first order equations yield different second order ones depending upon the assumptions made about the dependency of the unknown functions in the first order equations upon the functions which enter into the second order equations \cite{intrev}\cite{fai}.

\section{Lagrangian}
The construction of a Lagrangian for (\ref{complex2}) follows along the lines of \cite{fai}. Introduce an auxiliary field $\psi$ and consider the singular Lagrangian
\begin{equation}
{\cal L} \,=\, \left(\frac{\partial\bar\phi}{\partial t}\frac{\partial\psi}{\partial x}-\frac{\partial\psi}{\partial t}\frac{\partial\bar\phi}{\partial x}\right)\frac{\frac{\partial\phi}{\partial t}}{\frac{\partial\phi}{\partial x}}.\label{lag}
\end{equation}
The equation of motion corresponding to variations in the field $\psi$ is simply
equation (\ref{complex2}). Similarly for the variations with respect to $\bar\phi$ we obtain
\begin{equation}
\psi_x \phi_x\phi_{tt}-\psi_x \phi_t\phi_{tx}-\psi_t \phi_x\phi_{tx}+\psi_t\phi_t\phi_{xx}\,=\,0,\label{complex3}
\end{equation}
i.e. a similar equation with $\bar\phi$ replaced with $\psi$. But the third equation, corresponding to variations with respect to $\phi$ is just
\begin{equation}
\frac{\partial}{\partial t}\left[(\bar\phi_t\psi_x-\bar\phi_x\psi_t)\left(\frac{1}{\phi_x}\right)\right]
-\frac{\partial}{\partial x}\left[(\bar\phi_t\psi_x-\bar\phi_x\psi_t)\left(\frac{\phi_t}{\phi_x^2}\right)\right]\,=\,0.\label{complex4}
\end{equation}
This is satisfied if $\psi$ is a function of $\bar\phi$; then equation (\ref{complex3}) is the same as equation (\ref{complex2}). Incidentally, we see here a situation which has been remarked upon before in the context of free field equations \cite{dbf}, and equations arising from Born-Infeld Lagrangians, namely that the Lagrangian itself is a constant, or else a divergence on the space of solutions of the equations of motion. It is also evident that 
the factor $\displaystyle{\frac{\frac{\partial\phi}{\partial t}}{\frac{\partial\phi}{\partial x}}}$ may be replaced by any homogeneous function of $\displaystyle{\left(\frac{\partial\phi}{\partial t},\ \frac{\partial\phi}{\partial x}\right)}$ of weight zero, without affecting the equations of motion.
 
\section{Multi-field Lagrangian}
The Lagrangian can be constructed along similar lines to that for the single field;
one choice is
\begin{equation}
{\cal L}=\frac{\pd(\bar\phi^1,\ \bar\phi^2,\ \theta)}{\pd(x_1, x_2, x_3)}\left(\frac{\frac{\pd(\phi^1,\ \phi^2)}{\pd(x_1,\ x_2)}}{\frac{\pd(\phi^1,\ \phi^2)}{\pd(x_1,\ x_3)}}\right)\\,+\, {\rm cc}.\label{newlag}
\end{equation}
Variation with respect to $ \theta$ gives a combination of the equations of motion for $\phi^1$ and $\phi^2$ and their complex conjugates; variations with respect to the  fields
$\bar\phi^1$ and $\bar\phi^2$  yields other linear combinations which together imply the following equations,where $j$ takes the values (1,\ 2);

\begin{equation}
\det\left|\begin{array}{ccccc}
0&0&\bar\phi^1_{x_1}&\bar\phi^1_{x_2}& \bar\phi^1_{x_3} \\
0&0&\bar\phi^2_{x_1}&\bar\phi^2_{x_2}& \bar\phi^2_{x_3} \\
\phi^1_{x_1}&\phi^2_{x_1}&\phi^j_{x_1x_1}&\phi^j_{x_1x_2}&\phi^j_{x_1x_3}\\
\phi^1_{x_2}&\phi^2_{x_2}&\phi^j_{x_2x_1}&\phi^j_{x_2x_2}&\phi^j_{x_2x_3}\\
\phi^1_{x_3}&\phi^2_{x_3}&\phi^j_{x_3x_1}&\phi^j_{x_3x_2}&\phi^j_{x_3x_3}\end{array}
\right|\,=\,0.\label{result}
\end{equation}
and that $\theta$ is a function of $\bar\phi^1$ and $\bar\phi^2$, in much the same manner as $\psi$ is a function of $\bar\phi$ in the single field case.
As in the case of a single pair of complex fields, these equations follow from
a set of hydrodynamic equations.
\begin{eqnarray}
\frac{\pd u^i}{\pd x_1}&+& v^1\frac{\pd u^i}{\pd x_3}\,+\, v^2\frac{\pd u^i}{\pd x_2}\,=\,0\ \ i=1,2\label{hydro1}\\
\frac{\pd v^i}{\pd x_1}&+& u^1\frac{\pd v^i}{\pd x_3}\,+\, u^2\frac{\pd v^i}{\pd x_2}\,=\,0\ \ i=1,2.\label{hydro4}
\end{eqnarray} 
Once again, these equations remain the same up to a constant factor under a general linear transformation of co-ordinates; this may be seen most easily if they are written in a 
homogeneous notation by introducing vectors $\xi^\mu,\ \eta^\mu;\ \mu\,=\,0,1,2$ such that
$\displaystyle{u^i\, =\, \frac{\xi^i}{\xi^0},\ \ v^i\, =\, \frac{\eta^i}{\eta^0}}$ so that the equations may be written as
\[\sum_0^2 \xi^\mu\frac{\pd v^i}{\pd x_\mu}\,=\, 0;\ \ \ \sum_0^2 \eta^\mu\frac{\pd u^i}{\pd x_\mu}\,=\,0,\]
making the invariance up to a factor  of the hydrodynamic equations under linear transformations of the co-ordinates and the vectors $\vec\xi,\ \vec\eta$ manifest.

Set $u^i\,=\,\phi_i$ and choose the following set of these equations and their derivatives:
\begin{eqnarray}
&&\phi^1_{x_1}+v^1\phi^1_{x_2}+v^2\phi^1_{x_3}\,=\,0\nonumber\\
&&\phi^2_{x_1}+v^1\phi^2_{x_2}+v^2\phi^2_{x_3}\,=\,0\nonumber\\
&&\phi^1_{x_1x_1}+v^1\phi^1_{x_1x_2}+v^2\phi^1_{x_1x_3}+v^1_{x_1}\phi^1_{x_2}+v^2_{x_1}\phi^1_{x_3}\,=\,0\nonumber\\
&&\phi^1_{x_1x_2}+v^1\phi^1_{x_2x_2}+v^2\phi^1_{x_2x_3}+v^1_{x_2}\phi^1_{x_2}+v^2_{x_2}\phi^1_{x_3}\,=\,0\nonumber\\
&&\phi^1_{x_1x_3}+v^1\phi^1_{x_2x_3}+v^2\phi^1_{x_3x_3}+v^1_{x_3}\phi^1_{x_2}+v^2_{x_3}\phi^1_{x_3}\,=\,0.\nonumber
\end{eqnarray}
Eliminate the first derivatives ($\phi^1_{x_2},\ \phi^1_{x_3}$) from the final three equations ,
solve the first pair of equations for $(v^1,\ v^2)$ and subsitute in the undifferentiated terms of the result, setting $v^1 =\bar\phi^1,\ v^2=\bar\phi^2$. The ensuing equation is just  one member of (\ref{result}). 
\section{The fundamental hydrodynamic equations}
All second order integrable equations of the type discussed here and in earlier work
\cite{leznov},\cite{fai} are consequences of the general first order equations, for which an implicit solution may be constructed  following Leznov \cite{leznov2}.
Consider a $2n$ dimensional Euclidean space with indepenent co-ordinates $(x_i,\ \bar x_i,\ i=1\dots n)$ and construct the differential operators
\begin{equation}
D\,=\,\frac{\pd}{\pd x_n} +\sum^{n-1}_{j=1}u^j\frac{\pd}{\pd x_j},\ \ \ \
\bar D\,=\,\frac{\pd}{\pd\bar x_n} +\sum^{n-1}_{j=1}v^j\frac{\pd}{\pd\bar x_j}
\label{hoo}
\end{equation}
Since $D\bar x_i\,=\,0,\ \bar D x_i\,=\,0,$  $D$ may be considered a holomorphic differential operator and $\bar D$  an antiholomorphic operator.. Now imposing the zero curvature condition, $[D,\ \bar D]\,=\,0$ requires that
\begin{equation}
Dv^i\,\equiv\, v^i_{x_n}+\sum u^jv^i_{x_j}\,=\,0,\ \ \ \bar Du^i\,\equiv\, u^i_{\bar x_n}+\sum v^ju^i_{\bar x_j}\,=\,0.\label{lezimp}
\end{equation}
These are the general first order equations mentioned above. Since $D,\ \bar D$ commute, these equations imply that $\bar D v^i$ is a solution to the same equation as $v^i$ satisfies. The integration of the equation $D\bar D v^i\,=\,0$ requires that $\bar D v^i$
is a general anti-holomorphic function, hence,
\begin{equation}
\bar D v^i\,=\, v^i_{\bar x_n}+\sum v^jv^i_{\bar x_j}\,=\,V^i(v^k;\bar x_l),\ \ \  Du^i\,=\, u^i_{ x_n}+\sum u^ju^i_{\bar x_j}\,=\,U^i(u^k;x_l).\label{lezimp2}
\end{equation}
Indeed $f(\bar D)v^i$, for arbitrary differentiable $f$ is also a solution to the equation
for $v^i$.
Suppose now we take $(n-1)$ functions $\phi^i$ constrained by the $(n-1)$ relations
\begin{equation}
Q^i(\phi^j;x_k)\,=\,P^i(\phi^j;\bar x_k),\  i=1\dots n-1.\label{relate}
\end{equation}
The arbitrary functions $Q^i,\ P^i$ depend upon $(2n-1)$ co-ordinates.
They imply straightforwardly
\begin{equation}\phi^j_{x_k}\,=\, (P^i_{\phi^j}-Q^i_{\phi^j})^{-1}Q^i_{x_k},\
\phi^j_{\bar x_k}\,=\, -(P^i_{\phi^j}-Q^i_{\phi^j})^{-1}P^i_{\bar x_k}.\label{cons}
\end{equation}
Suppressing the vector indices, suppose $u$ is a function  $u(\phi, x)$  and
$v$ is a function  $u(\phi,\bar x)$.  Then the equations (\ref{lezimp}) imply that
\begin{equation}
D\phi^j\,=\,\phi^j_{x_n}+\sum_1^{n-1}v^k\phi_{x_k}\,=\,0,\\\
\bar D \phi^j\,=\,\phi^j_{\bar x_n}+\sum_1^{n-1}u^k\phi_{\bar x_k}\,=\,0,\label{lezimp3}
\end{equation}
In other words this requires that $\phi^j$ be both holomorphic and antiholomorphic in this definition of holomorphicity.
Subsituting the derivatives from (\ref{cons}) and multiplying on the left by the matrix $(P_\phi-Q_\phi)$ the equations become
\begin{equation}
Q^j_{x_n}+\sum_1^{n-1}v^kQ^j_{x_k}\,=\,0,\\\
P^j_{\bar x_n}+\sum_1^{n-1}u^kP^j_{\bar x_k}\,=\,0,\label{lezimp4}
\end{equation}
which leads to the identifications
\begin{equation}
v\,=\, -(Q_x)^{-1}Q_{x_n}, \ \ \ u\,=\, -(P_{\bar x})^{-1}P_{\bar x_n}.\label{defu}
\end{equation}
In consequence any function of $\phi,\bar x$ ($\phi, x$) will be annihilated by
$D\  (\bar D).$ In particular
\begin{equation}
D\phi\,=\,\bar D\phi \,=\,DQ\,=\,\bar DP\,=\,DQ\,=\,\bar D P\,=\,0.\label{holo}
\end{equation}
The last two results follow a forteriori from the equality $P\,=\,Q$.
The functions $\phi^j$ satisfy the multi-field complex Bateman equation
\cite{fai}.
\section{Partially integrable covariant equations.}
It appears likely that in the case where the difference between the number of dimensions of the base space exceeds that of the target space by more than one, the equations of motion are only partially integrable, though this is by no means a definitive conclusion.
In the case of one field dependent on three co-ordinates, the equation which results from the  Euclidean Lagrangian
\[ {\cal L} =\sqrt{\phi_{t}^2+ \phi_{x}^2+ \phi_{y}^2}\] 
is
\[
\phi_{tt}(\phi_{x}^2+ \phi_{y}^2)+\phi_{xx}(\phi_{y}^2+
\phi_{t}^2)+\phi_{yy}(\phi_{t}^2+ \phi_{x}^2)
=2\phi_{tx}\phi_t\phi_x+2\phi_{yt}\phi_y\phi_t+2\phi_{xy}\phi_x\phi_y.
\]

This equation possesses a large class of solutions given implicitly by solving
\[tF(\phi)+xG(\phi)+yK(\phi)\,=\,{\rm constant}\]
for $\phi$.
It comes from  the following first order system;
\begin{eqnarray}
uu_x+vv_x&=&u_t+v_y+v^2u_t-uv(u_y+v_t)+u^2v_y\nonumber\\
uv_x-vu_x &=& v_t-u_y,\nonumber
\end{eqnarray}
where $\displaystyle{u\,=\, \frac{\phi_t}{\phi_x},\ v\,=\, \frac{\phi_y}{\phi_x}}$.
This construction suggests further analysis to try to determine whether the system is fully integrable or not. It is also not known whether there exist other Lagrangian formulations of these equations. 

\section*{Acknowledgement}
The author is indebted to the Leverhulme Trust for the award of an Emeritus Fellowship and to the Clay Mathematics Institute for
employment during the investigations reported here. 

\end{document}